\documentclass[
aps,prd,
12pt,
nopreprintnumbers,
showpacs,
eqsecnum,
nofootinbib
]{revtex4-1}

\usepackage{graphicx}
\usepackage{amssymb}

\begin{document}

\title{Quantum BPS Cosmology}
\author{Nahomi Kan}\email[]{kan@gifu-nct.ac.jp}
\affiliation{National Institute of Technology, Gifu College,
Motosu-shi, Gifu 501-0495, Japan}
\author{Kiyoshi Shiraishi\footnote{Author to whom any
correspondence should be addressed.}}\email[]{shiraish@yamaguchi-u.ac.jp}
\affiliation{
Graduate School of Sciences and Technology for Innovation, Yamaguchi
University, Yamaguchi-shi, Yamaguchi 753--8512, Japan}
\author{Maki Takeuchi}\email[]{maki\_t@yamaguchi-u.ac.jp}
\affiliation{
Graduate School of Sciences and Technology for Innovation, Yamaguchi
University, Yamaguchi-shi, Yamaguchi 753--8512, Japan}
\author{Mai Yashiki}\email[]{yashikimi@nbu.ac.jp}
\affiliation{Faculty of School of Engineering, Nippon Bunri University, 
Oita-shi, Oita 870-0397, Japan}

\date{\today}

\begin{abstract}
There has been much discussion about the initial conditions
of the early Universe in the context of quantum theory. In this paper, we
construct the wave function and probability distribution by adopting the quantum
version of the BPS equation instead of the usual Wheeler--DeWitt equation in a
minisuperspace quantum cosmology with spatially uniform scalar fields. Although
the model treated in this study is technically valid for a limited form of scalar
potential, we show that it is possible to construct a conserved probability
current in our model. We also examine classical and quantum aspects of models
with the Dirac--Born--Infeld type scalar fields.
\end{abstract}

\maketitle

\section{Introduction}
\label{introduction}
Quantum cosmology \cite{Halliwell,Kiefer0,Kiefer1} in minisuperspace is based on
the idea that the Hamiltonian constraint in a gravitating system containing a
finite number of variables, such as scale factors and scalar fields, is
interpreted as a differential equation of the wave function of the Universe. This
equation, the Wheeler--DeWitt equation for a minisuperspace, becomes a hyperbolic
second-order partial differential equation by a straightforward construction.
This makes it difficult to define a positive-definite probability density.
Various solutions have been proposed for this problem in definition of the
probability density, including a simple and conventional adoption of the square of
the absolute value of the wave function. Furthermore, determining what kind of
``initial conditions'' to solve the Wheeler--DeWitt equation is the most
important when considering the very early universe. For this purpose, we have
only to choose plausible initial conditions to solve the equation.%
\footnote{Note that it is known that the wave
packet solution obtained by the conventional method and interpretation corresponds
to the classical solution in the gravitational system
\cite{KN,Kiefer2,Kiefer3,GS,KKST2}.}

 Recently, Russo \cite{Russo2,Russo1} has been studying a class of Einstein
gravity theories involving scalar fields in which the classical equations of
motion are reduced to first-order ordinary differential equations of time, such
as the BPS equations.%
\footnote{The same or similar systems have been considered by many authors
\cite{SB,BGLM,BLRR,SA,Lidsey,Taylor} from the past to the present, but
there has been little mutual citation between them.} 
The first-order differential equation
represents the streamline  in phase space, and has the advantage that the
behavior of the dynamical system can be directly obtained. Although solvable
systems are special cases, they often play an important role in the development
of all areas of physics. The solutions of BPS type equations may not, of course,
represent all possible solutions of the original equations of motion. Solution
with restrictions become effective only when the existence of underlying
symmetries, such as supersymmetry (or mock symmetry), is presumed. It seems that
the model including scalars discussed here includes the possibility of being
connected to fundamental theories with supersymmetry such as string theory.
Quantum cosmology with supersymmetry has already been actively researched
\cite{Graham1,Graham2,OSB,OPR,RB,Moniz}, but in this paper we will be careful to
analyze the model without specifying the symmetry.

 On the other hand, a quantum version of the BPS equation has already been used
through studies of partition functions in topological string theory
\cite{OVV,CV}. Therefore, we will use the quantum version of the BPS
equation as an alternative version of the Wheeler--DeWitt equation for the
similar model to Russo's model, and study the solution to the wave function of
the Universe obtained from the equation. This could provide a new perspective on
plausible choice of initial conditions for restricted wave functions of the
Universe. Our goal is to learn something useful about the probability
interpretation of such solutions.

 Therefore, although the origins of symmetries and restrictions are not
discussed in the current simple models, as mentioned earlier, we must choose a
solution that is compatible with the appropriate initial conditions of the
quantum Universe. We believe that there is some significance in examining whether
the wave function of the specious Universe can be obtained from the BPS-like
equation.

 The structure of this paper is as follows. Section \ref{sec2} reviews Russo's
work on classical cosmological equations. In Section \ref{sec3}, we consider the
BPS-like equations and their quantum version, and find the specific solution of
the wave function.  The
construction of probability density will be discussed in Section \ref{sec4}. 
In Section \ref{sec5}, we examine cosmology of the Dirac--Born--Infeld type model
by use of the BPS equation for the model. The final section provides a summary and
future prospects.

We use metric signature $(-+\cdots+)$ and units $16\pi G=c=\hbar=1$. 
$\mu, \nu,\dots=0, 1,\cdots, D-1$ are coordinate indices of spacetime, while
$i, j=1, 2,\cdots, D-1$ are indices for space.

\section{Review of classical cosmology with multiscalar fields}
\label{sec2}

In this section, we briefly review Russo's model \cite{Russo2,Russo1} in
$D$-dimensional spacetime. The action describing the system of gravitating
scalar fields is as follows:
\begin{equation}
S=\int d^Dx \sqrt{-g}\left[
R-\frac{1}{2}g^{\mu\nu}G_{ab}(\phi)\partial_\mu\phi^a
\partial_\nu\phi^b-V(\phi)
\right]\,,
\end{equation}
where $g$ denotes the determinant of the metric tensor $g_{\mu\nu}$,
$g^{\mu\nu}$ is the inverse of the metric tensor, $R$ is the scalar curvature, 
$\phi^a$ $(a=1,\cdots, N)$ are the scalar fields, and $V(\phi)$ is the
potential of scalar fields $\{\phi^a\}$. The metric of the internal space
$G_{ab}(\phi)$ is given by a symmetric matrix,
$G_{ab}=G_{ba}$ and is generally dependent on scalar fields $\{\phi^a\}$.
Further, the metric of the $D$ dimensional spacetime is assumed as
\begin{equation}
ds^2=g_{\mu\nu}dx^\mu dx^\nu=-e^{2\gamma
\varphi(t)}dt^2+e^{2\beta\varphi(t)}\sum_{i=1}^{D-1} (dx^i)^2
=-e^{2\gamma
\varphi(t)}dt^2+a^2(t)\sum_{i=1}^{D-1} (dx^i)^2\,,
\label{metric}
\end{equation}
where $\gamma$ is a constant, and
\begin{equation}
\beta=\left(\sqrt{2(D-1)(D-2)}\right)^{-1}\,,
\end{equation}
and $a(t)$ is a scale factor of the flat, homogeneous and isotropic space.
Note that this metric becomes the standard
Friedmann--Lema\^{\i}tre--Robertson--Walker metric when $\gamma=0$, while the
metric becomes a conformally-flat metric when $\gamma=\beta$.

Assuming that all scalar fields are also spatially uniform, that is, a
function of time only, $\phi^a=\phi^a(t)$,
the effective Lagrangian on these variables is given by\footnote{The
Gibbons--Hawking--York term \cite{York,GH} is implicitly taken into account, as a
standard approach in Einstein gravity.}
\begin{equation}
L=-\frac{1}{2}e^{\delta\varphi}\dot{\varphi}^2
+\frac{1}{2}e^{\delta\varphi}G_{ab}(\phi)
\dot{\phi}^a\dot{\phi}^b
-e^{(2\gamma+\delta)\varphi}V(\phi)\,,
\label{Lag0}
\end{equation}
where $\delta=(D-1)\beta-\gamma$ and 
the dot $(\dot{~})$ denotes the time derivative.

The equations of motion derived from the Lagrangian (\ref{Lag0}) are
\begin{equation}
\frac{d}{dt}
\left[e^{\delta\varphi}
G_{ab}\dot{\phi}^b\right]=-e^{(2\gamma+\delta)\varphi}\partial_aV
+\frac{e^{\delta\varphi}}{2}(\partial_aG_{bc})
\dot{\phi}^b\dot{\phi}^c\,,
\label{eomp0}
\end{equation}
and
\begin{equation}
\frac{d}{dt}(-e^{\delta\varphi}\dot{\varphi})
=-\frac{1}{2}\delta\, e^{\delta\varphi}\dot{\varphi}^2
+\frac{1}{2}\delta\, e^{\delta\varphi}G_{cd}
\dot{\phi}^c\dot{\phi}^d-(2\gamma+\delta)
e^{(2\gamma+\delta)\varphi}V\,,
\label{eomv0}
\end{equation}
where $\partial_a=\frac{\partial}{\partial\phi^a}$.

These equations are satisfied by the following two simultaneous equations
including the first-order derivative:
\begin{equation}
e^{\delta\varphi}\dot{\varphi}=\epsilon\partial_\varphi\mathcal{W}\,,\quad
e^{\delta\varphi}
\dot{\phi}^a=-\epsilon G^{ab}\partial_b\mathcal{W}\,,
\label{BPS}
\end{equation}
where the constant $\epsilon$ satisfies $\epsilon^2=1$, 
$\partial_\varphi=\frac{\partial}{\partial\varphi}$, and $G^{ab}$ is the
inverse matrix of
$G_{ab}$,  if the scalar potential takes the following form:
\begin{eqnarray}
V(\phi)&=&e^{-2(\gamma+\delta)\varphi}\left[\frac{1}{2}(\partial_\varphi\mathcal{W})^2
-\frac{1}{2}G^{ab}(\phi)\partial_a
\mathcal{W}\partial_b\mathcal{W}\right]\nonumber \\
&=&\frac{1}{2}\alpha^2({W(\phi}))^2
-\frac{1}{2}G^{ab}(\phi)\partial_a
{W(\phi)}\partial_b{W(\phi)}\,,
\end{eqnarray}
where
\begin{equation}
\mathcal{W}=e^{(\gamma+\delta)\varphi}W(\phi)=e^{(D-1)\beta\varphi}W(\phi)
=e^{\alpha\varphi}W(\phi)=a^{D-1}W(\phi)\,,
\end{equation}
and
\begin{equation}
\alpha=(D-1)\beta=\sqrt{\frac{D-1}{2(D-2)}}\,.
\end{equation}
The potential $V(\phi)$ is mostly determined by $W(\phi)$, so we call $W(\phi)$
prepotential.

We find that the Lagrangian (\ref{Lag0}) can be rewritten in the form
\begin{eqnarray}
L&=&-\frac{1}{2}e^{-\delta\varphi}(e^{\delta\varphi}\dot{\varphi}-\epsilon
\partial_\varphi\mathcal{W})^2+\frac{1}{2}e^{-\delta\varphi}
G_{cd}\left[e^{\delta\varphi}\dot{\phi}^c
+\epsilon G^{ce}\partial_e\mathcal{W}\right]
\left[e^{\delta\varphi}\dot{\phi}^d
+\epsilon G^{de}\partial_e\mathcal{W}\right]\nonumber \\
& &-\epsilon(\dot{\varphi}\partial_\varphi\mathcal{W}+
\dot{\phi}^a\partial_a\mathcal{W})\,,
\end{eqnarray}
so, it is apparent that the equations (\ref{BPS}) represent a stationary point of
the action $S=\int L dt$.

Based on the Lagrangian (\ref{Lag0}), the conjugate momenta of $\varphi$ and
$\phi^a$ are expressed by
\begin{equation}
\Pi_\varphi=\frac{\partial
L}{\partial\dot{\varphi}}=-e^{\delta\varphi}\dot{\varphi}\,,\quad
\Pi_a=\frac{\partial L}{\partial\dot{\phi}^a}=e^{\delta\varphi}G_{ab}(\phi)
\dot{\phi}^b\,,
\label{pv}
\end{equation}
respectively,
and the Hamiltonian is expressed by 
\begin{equation}
\mathcal{H}=-\frac{1}{2}e^{-\delta\varphi}\Pi_\varphi^2
+\frac{1}{2}e^{-\delta\varphi}G^{ab}(\phi)\Pi_a
\Pi_b+e^{(2\alpha-\delta)\varphi}V(\phi)\,.
\end{equation}
Therefore, if the BPS-like equations (\ref{BPS}) are substituted in the
Hamiltonian, we find that $\mathcal{H}=0$ holds classically.
The Hamiltonian constraint is automatically satisfied for the solutions of
(\ref{BPS}) in this case as a gravitating system.%
\footnote{As is well known, the Hamiltonian constraint is derived by,
replacing $t\rightarrow Nt$ in the action $S=\int L dt$ and regarding that the
variation
$\frac{\delta S}{\delta N}$ vanishes \cite{Halliwell,Kiefer0,Kiefer1}.}

The BPS-like equations (\ref{BPS}) can directly represent the flow of solutions
in phase space, so they are very convenient for examining the relationship
between initial conditions and subsequent evolution.

\section{Quantum version of BPS equations and the solution}
\label{sec3}

According to the paper \cite{OVV}, the BPS equation can be utilized for
quantization. Letting the wave function be $\Phi(\varphi,\phi)$, the quantization
is given by substitution of momenta
\begin{equation}
\Pi_\varphi\rightarrow -i\frac{\partial}{\partial\varphi}\,,\quad
\Pi_a\rightarrow -i\frac{\partial}{\partial\phi^a}\,,
\label{rm}
\end{equation}
as operators on the wave function \cite{Halliwell,Kiefer0,Kiefer1}, and then, the
resulting equations from (\ref{BPS}) are
\begin{equation}
\left[-i\partial_\varphi+\epsilon\partial_\varphi\mathcal{W}\right]
\Phi(\varphi,\phi)=0\,,\quad
\left[-i\partial_a+\epsilon\partial_a\mathcal{W}\right]
\Phi(\varphi,\phi)=0\,.
\end{equation}
As a simple solution to these equations,
\begin{equation}
\Phi(\varphi,\phi)=e^{-i\epsilon\mathcal{W}(\varphi,\phi)}\,,
\label{wkbe}
\end{equation}
can be considered. 

Let us take a look at the differences between the commonly used
Wheeler--DeWitt equation and the quantum BPS equations. 
The Wheeler--DeWitt equation is derived by replacing (\ref{rm}) in the
Hamiltonian constraint $\mathcal{H}=0$, and can be read as
\begin{equation}
\left[\partial_\varphi^2-G^{ab}(\phi)\partial_a
\partial_b+2e^{2\alpha\varphi}V(\phi)\right]\Phi=0\,,
\label{WDW0}
\end{equation}
up to operator orderings. Note that the Wheeler--DeWitt equation is independent
of the choice of the constant $\delta$, in turn the choice of $\gamma$.

As is clear from the form of the solution (\ref{wkbe}) of the BPS equations, this
solution is a solution of the Hamilton--Jacobi equation for the Wheeler--DeWitt
equation (\ref{WDW0}),
\begin{equation}
\Phi=e^{iS}\,,\quad
-(\partial_\varphi S)^2+G^{ab}(\phi)\partial_a S
\partial_b S+2e^{2\alpha\varphi}V(\phi)=0\,,
\label{HJ}
\end{equation}
and its simple solution is $S=-\epsilon\mathcal{W}$.
Therefore, the semiclassical wave function (or the WKB wave
function) (\ref{wkbe}) corresponds to a specific solution representing a certain
classical trajectory.
 By the way, as a general
matter, it is natural that the difference obtained by quantizing classical
equations can always appear, because there are ambiguities which depends on the
operator-ordering prescription.%
\footnote{A similar attitude has been adopted in work on minisuperspace
quantum cosmology from the early-days study \cite{Halliwell} to the most recent
study
\cite{Godet}.}
 The discrepancy from the specific ordering
may modify the solution for the wave function multiplicatively, so we can neglect
the possible correction which could also come from the one-loop effect of the
same order of the Planck constant $\hbar$.

In general, the wave function of the Universe, which is given as the solution of
the Wheeler--DeWitt equation, is written by an ensemble of semi-classical
universes
\cite{D37,D50}
\begin{equation}
\Phi=\sum_k e^{iS_k(c)}\chi_k(c,q)\,,
\end{equation}
where the sum is taken over the classical trajectory, $S_k(c)$ is the classical
solution of the Hamilton--Jacobi equation, and $\chi_k(c,q)$ is the contribution
of the quantum variables $q$. 
Thus, taking only
$S=-\epsilon\mathcal{W}$ as the solution corresponds to choosing a particular
classical solution for the wave function.

The choice of Hawking's no-boundary condition \cite{HH,Hawking} or Vilenkin's
tunneling wave function \cite{D33,D37} are  conjectures that attempt to select a
solution representing the initial state of the Universe among the general
solutions. By specifying the BPS solution, we are proposing a new conjecture for
the problem of the initial condition regarding the wave function of the Universe
in minisuperspace quantum cosmology. It would be nice if it could be connected to
the classical inflationary solution for the Universe.

As the reader has already noticed, in the system we are currently dealing with,
we are considering a flat universe, which naturally requires a different treatment
from the quantum cosmology represented by Hawking and Vilenkin in the models with
a homogeneous space of positive curvature. We will explain how to think about wave
functions in our system and the case of specific potentials $V(\phi)$ and
prepotentials
$W(\phi)$ in later sections.

\section{Construction of probability density}
\label{sec4}

For an interpretation of the wave function, we would like to obtain
conserved quantity for constructing probability density.
Taking into account the covariance of the minisuperspace, i.e., by the so-called
Laplace--Beltrami ordering,  we write down a more
sophisticated Wheeler--DeWitt equation
\begin{equation}
\left[e^{-p\varphi}\partial_\varphi
e^{p\varphi}\partial_\varphi-\frac{1}{\sqrt{G}}\partial_a\sqrt{G}G^{ab}
\partial_b+2e^{2\alpha\varphi}V(\phi)\right]\Phi=0\,,
\end{equation}
where $G\equiv \det G_{ab}$, and the constant $p$ represents an ordering
ambiguity.

The conservation equation created from the second-order partial differential
equation \cite{D33,D37,D39,HGC},
\begin{equation}
\partial_\varphi\rho+\nabla_a J^a=0\,,
\end{equation}
where
\begin{equation}
\rho\equiv
\frac{i}{2}\epsilon
e^{p\varphi}(\Phi^*\partial_\varphi\Phi-\partial_\varphi\Phi^*\Phi)\,,\quad
J^a\equiv -\frac{i}{2}\epsilon
e^{p\varphi}(\Phi^*\nabla^a\Phi-\nabla^a\Phi^*\Phi)\,,
\end{equation}
where $\nabla_a$ is a covariant derivative in metric $G_{ab}$.
Thus, we conclude that
\begin{equation}
\frac{d}{d\varphi}\int \rho\,\sqrt{G}\,d^N\phi=0\,,
\end{equation}
where $d^N\phi\equiv \prod_{a=1}^Nd\phi^a$, and the probability measure can be
taken as \cite{D33,D37,D39,HGC}
\begin{equation}
dP=\rho\sqrt{G}\,d^N\phi\,.
\end{equation}

Now, substituting the solution $\Phi=e^{-i\epsilon\mathcal{W}}$ into 
the above definitions, we find that
\begin{equation}
\alpha
e^{p\varphi}\sqrt{G(\phi)}\,\mathcal{W}=\alpha
e^{(\alpha+p)\varphi}\sqrt{G(\phi)}\,W(\phi)
\end{equation}
expresses the probability density for $\{\phi^a\}$ at a fixed $\varphi$.
As is often thought, $\varphi$ is defined as time, or the increase of a volume
element
$a^{D-1}=e^{\alpha\varphi}$ indicates time development of the Universe.
Furthermore, if the constant $p$ takes the value $-\alpha$,
the probability density is proportional to
\begin{equation}
\sqrt{G(\phi)}\,W(\phi)\,,
\end{equation}
which is independent of $\varphi$.

This becomes a positive definite value only if the prepotential $W(\phi)$ is
chosen as a positive-definite function. That is, the resolution of the problem of
the positivity of the probability measure is set aside in the present analysis.%
\footnote{Note that the change the sign of $\epsilon$ is equivalent to the
change $W\rightarrow-W$. } 

We propose that the initial value for scalar fields are determined by the maximum
of the positive prepotential $W(\phi)$,%
\footnote{Note that this means the maximum expansion rate, since
$\frac{\dot{a}}{a}$  is proportional to $W(\phi)$, which is seen from 
(\ref{BPS}).} whereas nothing is said about the initial condition on the scale of
the Universe. This is natural in a sense, since in the present analysis, the
volume element is treated as time. Also, since we are dealing with a model in
which the curvature of the space at any time slice is zero, the singularity
problem does not appear seriously. 
We also pay attention to the fact that the classical action
becomes $\int_{t_i}^{t_f} L dt=-\epsilon(\mathcal{W}|_{t_f}-\mathcal{W}|_{t_i})$,
if the equations (\ref{BPS}) hold. Therefore, the extremum of the classical action
corresponds to the maximum probability density. In this case, the factor
$\sqrt{G}$ simply arises from  the Jacobian of the scalars.
 
Here, we give an example with a single scalar field (i.e., $G_{ab}\rightarrow 1$).
Suppose that
\begin{equation}
W(\phi)=w^2(\cosh m\phi)^{-b}\,,
\end{equation}
where the constants $w$, $m$ and $b$ takes positive values.
In this case, the most probable value of $\phi$ in the quantum Universe is
near $\phi=0$.

The classical trajectory from (\ref{BPS}) satisfies
\begin{equation}
\frac{d\phi}{d\varphi}=\frac{\dot{\phi}}{\dot{\varphi}}=-\frac{W'(\phi)}{\alpha
W(\phi)}=\frac{bm}{\alpha}\tanh m\phi\,,
\end{equation}
where $W'(\phi)=\frac{dW}{d\phi}$.
The solution of this equation is found to be
\begin{equation}
e^{\alpha\varphi}=a^{D-1}=C (\sinh m\phi)^{\frac{\alpha^2}{bm^2}}\,,
\end{equation}
where $C$ is a constant.%
\footnote{The constant $C$ is not physical, since it can be absorbed into the
rescaling of spatial coordinates.} One can see that the scale factor becomes
$a\approx 0$ when
$\phi\rightarrow 0$.
So, in this case, the BPS equations also give the initial condition $a\approx 0$.

We solve the BPS-like equations (\ref{BPS}), using the standard cosmic time in the
Friedmann--Lema\^{\i}tre--Robertson--Walker metric, obtained by choosing
$\gamma=0\, (\leftrightarrow \delta=\alpha)$ in the metric (\ref{metric}). If
$\gamma=0$, the equations (\ref{BPS}) for a single scalar field becomes
\begin{equation}
\dot{\varphi}=\alpha W(\phi)\,,\quad \dot{\phi}=-W'(\phi)\,,
\end{equation}
and then, the expansion rate is
\begin{equation}
H=\frac{\dot{a}}{a}=\alpha\beta W(\phi)\,.
\end{equation}

In the early time in the Universe, when $m\phi\ll 1$, we find
\begin{equation}
e^{\alpha\varphi}=a^{D-1}\approx Ce^{\alpha^2w^2(t-t_0)}\,,\quad
\phi\approx m^{-1} e^{b\,m^2w^2(t-t_0)}\,,
\end{equation}
where $t_0$ is a constant.
This shows the de Sitter expansion, of which expansion rate is
$H=\frac{\alpha^2w^2}{D-1}=\frac{w^2}{2(D-2)}$.

In the late time, when $m\phi\gg 1$, we find
\begin{equation}
e^{\alpha\varphi}=a^{D-1}\approx
C[b^2m^2w^2(t-t_1)]^{\frac{\alpha^2}{b^2m^2}}\,,\quad
\phi\approx \frac{1}{bm}\ln[2^bb^2m^2w^2(t-t_1)]\,,
\label{as}
\end{equation}
where $t_1$ is a constant.
For $m\phi\gg 1$, the potential takes the asymptotic form
\begin{equation}
V(\phi)=\frac{1}{2}\frac{w^4}{(\cosh
m\phi)^{2b}}\left(\alpha^2-b^2m^2\tanh^2
m\phi\right)\approx\frac{1}{2}2^{2b}w^4(\alpha^2-b^2m^2)e^{-2bm\phi}
\quad (\mbox{for~}b^2m^2\ne \alpha^2)\,.
\end{equation}
The cosmological model of a scalar field with an exponential
potential has been studied by many authors until \cite{Russo0,Neupane,ENO}. 
They found that the potential $e^{-\lambda\phi}$ leads to $e^{\alpha\varphi}\sim
t^{4\alpha^2/\lambda^2}$ and $\phi\sim\frac{2}{\lambda}\ln t$.
The asymptotic solution (\ref{as}) coincides with their solution. 
Interestingly, the positivity of the potential at $m\phi\gg 1$ corresponds to a
condition of accelerating expansion, $\frac{\alpha^2}{b^2m^2}>1$, i.e. the
power-law inflation. In an exceptional case with $b^2m^2=\alpha^2$, we find
\begin{equation}
V(\phi)=\frac{1}{2}\frac{\alpha^2w^4}{(\cosh
m\phi)^{2b+2}}\approx\frac{1}{2}2^{2b+2}\alpha^2w^4e^{-2(b+1)m\phi}
\quad (\mbox{for~}b^2m^2=\alpha^2)\,,
\end{equation}
asymptotically. In this case, the relation between the coefficient of the exponent
of the asymptotic potential and the exponent in the scale factor differs from
that in \cite{Russo0,Neupane,ENO}.


Incidentally, the slow-roll parameters in \cite{Russo2} are expressed by the
prepotential $W(\phi)$, as
\begin{eqnarray}
\varepsilon_H&=&-\frac{\dot{H}}{H^2}=\frac{\dot{\phi}^2}{2(D-2)H^2}=2(D-2)\frac{W'(\phi)^2}{W(\phi)^2}\,,\quad
\\
\eta_H&=&-\frac{\ddot{H}}{2H\dot{H}}=-\frac{\ddot{\phi}}{H\dot{\phi}}
=2(D-2)\frac{W''(\phi)}{W(\phi)}\,.
\end{eqnarray}
For the present example, $W(\phi)=w^2[\cosh(m\phi)]^{-b}$, we find
\begin{equation}
\varepsilon_H=2(D-2)b^2m^2\tanh^2m\phi\,,\quad
\eta_H=2(D-2)bm^2[(b+1)\tanh^2m\phi-1]\,.
\end{equation}
Thus, the slow-roll condition $\varepsilon_H\ll 1$ is satisfied if
$m\phi\ll 1$ and/or $bm\ll 1$.

In summary, this example shows that the Universe could experience the nearly de
Sitter phase and the subsequent power-law inflation phase.

A comment should be placed here on the ``classicalization''.
The initial fluctuation of the scalar field and the scale factor in quantum
cosmology should produce corresponding classical quantities which obey the
classical equation of motion in classical cosmology. 
The validity of this classicalization needs to be explained theoretically.
Here, we consider the model with the simplest assumption that the initial state
is described by $\phi\approx 0$ and $a\approx 0$, including fluctuations, and
that the system rapidly transitions to classical system.


Finally in this section, we would like to give a brief comment on the Euclidean
treatment of the BPS equations. The authors of \cite{OVV} suggested the use of
the imaginary time and found the Hartle--Hawking state in their theory.
The use of the imaginary time yields
\begin{equation}
\Phi\propto e^{\mp\mathcal{W}}\,.
\end{equation}
In this case, we should consider the commonly used probability density,
$|\Phi|^2$. Also in this case, the extremum of the prepotential $W(\phi)$ gives
the most probable initial value of the scalar fields.

\section{Dirac--Born--Infeld type model}
\label{sec5}

The BPS-like equation for gravitating Dirac--Born--Infeld type scalar field
theory was found in \cite{ST,AST}. The action for the gravitating 
Dirac--Born--Infeld scalar fields is
\begin{eqnarray}
S&=&\int d^Dx \sqrt{-g}\left[
R-\frac{1}{f(\phi)}\sqrt{1+f(\phi)g^{\mu\nu}G_{ab}(\phi)\partial_\mu\phi^a
\partial_\nu\phi^b}-V(\phi)
\right]\nonumber \\
&=&\int d^Dx \sqrt{-g}\left[
R-\frac{1}{f(\phi)}\left\{\sqrt{1+f(\phi)g^{\mu\nu}G_{ab}(\phi)\partial_\mu\phi^a
\partial_\nu\phi^b}-1\right\}-U(\phi)
\right]\,,
\end{eqnarray}
where $f(\phi)$ is a function of the scalar fields $\{\phi^a\}$ and
$U(\phi)=V(\phi)+\frac{1}{f(\phi)}$.
Notice that the kinetic term reduces to the one in the previous canonical scalar
model in the limit of $f\rightarrow 0$. Using the same ansatze for the metric and
scalar fields as in Section
\ref{sec2}, we find the effective Lagrangian
\begin{equation}
L=-\frac{1}{2}e^{\delta\varphi}\dot{\varphi}^2
-\frac{e^{(\delta+2\gamma)\varphi}}{f(\phi)}
\sqrt{1-{e^{-2\gamma\varphi}}f(\phi)G_{ab}(\phi)
\dot{\phi}^a\dot{\phi}^b}
-e^{(\delta+2\gamma)\varphi}V(\phi)\,.
\label{DBIL}
\end{equation}
The equations of motion obtained from this Lagrangian are
\begin{eqnarray}
& &\frac{d}{dt}
\left[\left(\sqrt{1-{e^{-2\gamma\varphi}}fG_{cd}
\dot{\phi}^c\dot{\phi}^d}\right)^{-1}e^{\delta\varphi}
G_{ab}\dot{\phi}^b\right]\nonumber \\
&=&-e^{(2\gamma+\delta)\varphi}\partial_aV
+\frac{e^{\delta\varphi}}{2}\left(\sqrt{1-{e^{-2\gamma\varphi}}fG_{cd}
\dot{\phi}^c\dot{\phi}^d}\right)^{-1}(\partial_aG_{be})
\dot{\phi}^b\dot{\phi}^e\nonumber \\
& &
+e^{(2\gamma+\delta)\varphi}\frac{\partial_af}{2f^2}\left[
\sqrt{1-{e^{-2\gamma\varphi}}fG_{cd}
\dot{\phi}^c\dot{\phi}^d}+\left(\sqrt{1-{e^{-2\gamma\varphi}}fG_{cd}
\dot{\phi}^c\dot{\phi}^d}\right)^{-1}\right]\,,
\label{eomp}
\end{eqnarray}
\begin{eqnarray}
\frac{d}{dt}(-e^{\delta\varphi}\dot{\varphi})
&=&-\frac{\delta}{2}e^{\delta\varphi}\dot{\varphi}^2
-(\gamma+\delta)\frac{e^{(2\gamma+\delta)\varphi}}{f}
\sqrt{1-{e^{-2\gamma\varphi}}fG_{cd}
\dot{\phi}^c\dot{\phi}^d}\nonumber \\
& &-\gamma \frac{e^{(2\gamma+\delta)\varphi}}{f}
\left(\sqrt{1-{e^{-2\gamma\varphi}}fG_{cd}
\dot{\phi}^c\dot{\phi}^d}\right)^{-1}-(2\gamma+\delta)
e^{(2\gamma+\delta)\varphi}V\,.
\label{eomv}
\end{eqnarray}
These equations are solved by
\begin{equation}
e^{\delta\varphi}\dot{\varphi}=\epsilon\partial_\varphi\mathcal{W}\,,\quad
\left(\sqrt{1-{e^{-2\gamma\varphi}}fG_{cd}
\dot{\phi}^c\dot{\phi}^d}\right)^{-1}e^{\delta\varphi}
\dot{\phi}^a=-\epsilon G^{ab}\partial_b\mathcal{W}\,,
\label{bps2}
\end{equation}
provided that
\begin{eqnarray}
V(\phi)&=&\frac{1}{2}e^{-2\alpha\varphi}(\partial_\varphi\mathcal{W})^2
-\frac{1}{f}\sqrt{1+{e^{-2\alpha\varphi}}fG^{ab}\partial_a
\mathcal{W}\partial_b\mathcal{W}}\nonumber \\
&=&\frac{1}{2}\alpha^2({W(\phi}))^2
-\frac{1}{f(\phi)}\sqrt{1+f(\phi)G^{ab}(\phi)\partial_a
{W(\phi)}\partial_b{W(\phi)}}\,,
\end{eqnarray}
where
\begin{equation}
\mathcal{W}=e^{\alpha\varphi}W(\phi)=e^{(D-1)\beta\varphi}W(\phi)\,.
\end{equation}
Note that
\begin{equation}
\sqrt{1-{e^{-2\gamma\varphi}}fG_{ab}
\dot{\phi}^a\dot{\phi}^b}\sqrt{1+{e^{2\alpha\varphi}}fG^{ab}\partial_a
\mathcal{W}\partial_b\mathcal{W}}=1\,,
\label{id}
\end{equation}
if the equations (\ref{bps2}) hold,
and notice that $\partial_a G^{bc}=-G^{bd}G^{ce}\partial_aG_{de}$.

Then, the Lagrangian (\ref{DBIL}) can be rewritten by
\begin{eqnarray}
L&=&-\frac{1}{2}e^{-\delta\varphi}(e^{\delta\varphi}\dot{\varphi}-\epsilon
\partial_\varphi\mathcal{W})^2\nonumber \\
& &+\frac{1}{2}e^{-\delta\varphi}\sqrt{1-{e^{-2\gamma\varphi}}fG_{ab}
\dot{\phi}^a\dot{\phi}^b}\,G_{cd}\left[\left(\sqrt{1-{e^{-2\gamma\varphi}}fG_{ab}
\dot{\phi}^a\dot{\phi}^b}\right)^{-1}e^{\delta\varphi}\dot{\phi}^c
+\epsilon G^{ce}\partial_e\mathcal{W}\right]\nonumber \\
& &\qquad\qquad\qquad\times
\left[\left(\sqrt{1-{e^{-2\gamma\varphi}}fG_{ab}
\dot{\phi}^a\dot{\phi}^b}\right)^{-1}e^{\delta\varphi}\dot{\phi}^d
+\epsilon G^{de}\partial_e\mathcal{W}\right]\nonumber \\
& &-\frac{1}{2}\frac{e^{(2\gamma+\delta)\varphi}}{f}
\left(\sqrt{1-{e^{-2\gamma\varphi}}fG_{ab}
\dot{\phi}^a\dot{\phi}^b}\right)^{-1}\nonumber \\
& &\qquad\qquad\qquad\times\left[
\sqrt{1-{e^{-2\gamma\varphi}}fG_{ab}
\dot{\phi}^a\dot{\phi}^b}\sqrt{1+{e^{-2\alpha\varphi}}fG^{ab}\partial_a
\mathcal{W}\partial_b\mathcal{W}}-1\right]^2\nonumber \\
& &-\epsilon(\dot{\varphi}\partial_\varphi\mathcal{W}+
\dot{\phi}^a\partial_a\mathcal{W})\,.
\end{eqnarray}
Obviously, the limit $f\rightarrow 0$ yields the Russo's canonical model reviewed
in Section \ref{sec2}.

The conjugate momenta in the present system are
\begin{equation}
\Pi_\varphi=-e^{\delta\varphi}\dot{\varphi}\,,\quad
\Pi_a=\frac{e^{\delta\varphi}G_{ab}\dot{\phi}^b}{\sqrt{
1-{e^{-2\gamma\varphi}}fG_{cd}\dot{\phi}^c\dot{\phi}^d}}\,,
\label{mom2}
\end{equation}
and then the Hamiltonian can be found as
\begin{equation}
\mathcal{H}=-\frac{1}{2}e^{-\delta\varphi}\Pi_\varphi^2
+\frac{e^{(2\alpha-\delta)\varphi}}{f}
\sqrt{1+{e^{-2\alpha\varphi}}fG^{ab}\Pi_a
\Pi_b}+e^{(2\alpha-\delta)\varphi}V(\phi)\,.
\end{equation}
The classical Hamiltonian constraint $\mathcal{H}=0$ is satisfied by taking
(\ref{bps2}) with (\ref{mom2}).

The Wheeler--DeWitt equation $\mathcal{H}\Phi=0$ is obtained by replacing the
momenta with the operators.
The resulting equation is highly nonlinear differential equation in the present
case. Nevertheless, the Hamilton--Jacobi equation obtained by setting
$\Phi=e^{iS}$ and assuming the WKB approximation
$|\partial_a\partial_bS|\ll|\partial_aS|^2$ becomes
\begin{equation}
-\frac{1}{2}(\partial_\varphi S)^2
+\frac{e^{2\alpha\varphi}}{f}
\sqrt{1+{e^{-2\alpha\varphi}}fG^{ab}\partial_a S
\partial_bS}+e^{2\alpha\varphi}V(\phi)=0\,.
\label{hj2}
\end{equation}
Thus, the solution of the quantum BPS equations
\begin{equation}
\left[-i\partial_\varphi+\epsilon\partial_\varphi\mathcal{W}\right]
\Phi(\varphi,\phi)=0\,,\quad
\left[-i\partial_a+\epsilon\partial_a\mathcal{W}\right]
\Phi(\varphi,\phi)=0\,,
\end{equation}
which is the same as in the Section \ref{sec2}, that is,
\begin{equation}
\Phi=e^{-i\epsilon\mathcal{W}}\,,
\end{equation}
is also the solution of the Hamilton--Jacobi equation (\ref{hj2}).

At last, unlike for the model in Section \ref{sec2}, it is difficult to write
down the exact conservation equation with the present wave function.
However, under the WKB approximation already adopted, the probability density
should take the same form,
\begin{equation}
\sqrt{G(\phi)}\,W(\phi)\,.
\end{equation}
Incidentally, the Wheeler--DeWitt equation for a similar model was studied in
\cite{Psinas}, which also included an analysis of the lowest order of scalar
derivatives.

That aside, we note that we find the same probability measure in the
Dirac--Born--Infeld type model as in the canonical scalar model, which is
recovered by sending $f$ to zero.
Further, it has been noted that the classical value of
$e^{iS}$ is proportional to $e^{-i\epsilon\mathcal{W}}$, regardless of the
function $f$. Thus, as in the canonical case, with $f\rightarrow 0$, the maximum
probability density is taken to be given by the extremum of the prepotential.

Now, let us consider the similar example as in the previous section. 
Apparently, the de Sitter phase and the power-law behavior can be found by
considering the asymptotic form of $W(\phi)$ and for an appropriate function
$f(\phi)$.
In the following, for simplicity, we here assume $b=1$ and $f$ takes a constant
value. That is,
\begin{equation}
W(\phi)=w^2(\cosh m\phi)^{-1}\,.
\label{kk}
\end{equation}
The equations (\ref{bps2}) and (\ref{id}) lead to
\begin{equation}
\sqrt{1+f (W'(\phi))^2}\frac{d\phi}{d\varphi}=-\frac{W'(\phi)}{\alpha
W(\phi)}\,,
\end{equation}
and substitution of (\ref{kk}) yields
\begin{equation}
\sqrt{1+fm^2w^2 \frac{\sinh^2 m\phi}{\cosh^4
m\phi}}\frac{d\phi}{d\varphi}=\frac{m}{\alpha}\tanh m\phi\,.
\end{equation}
The solution of this equation is
\begin{eqnarray}
e^{\alpha\varphi}&=&C(\sinh m\phi)^\frac{\alpha^2}{m^2}\nonumber \\
&
&\cdot\left[\frac{2\cosh^2m\phi+fm^2w^2+2\sqrt{\cosh^4m\phi+fm^2w^2\sinh^2m\phi}}
{2\cosh^2m\phi+fm^2w^2\sinh^2m\phi+2\sqrt{\cosh^4m\phi+fm^2w^2\sinh^2m\phi}}\right]
^\frac{\alpha^2}{2m^2}\nonumber \\ &
&\cdot\exp\left[-\frac{\sqrt{fm^2w^2}\alpha^2}{2m^2}\arctan\frac{\sqrt{fm^2w^2}(1-\sinh^2m\phi)}
{2\sqrt{\cosh^4m\phi+fm^2w^2\sinh^2m\phi}}\right]\quad (\mbox{for~}f>0)
\,,
\end{eqnarray}
\begin{eqnarray}
e^{\alpha\varphi}&=&C(\sinh m\phi)^\frac{\alpha^2}{m^2}\nonumber \\
&
&\cdot\left[\frac{2\cosh^2m\phi+fm^2w^2+2\sqrt{\cosh^4m\phi+fm^2w^2\sinh^2m\phi}}
{2\cosh^2m\phi+fm^2w^2\sinh^2m\phi+2\sqrt{\cosh^4m\phi+fm^2w^2\sinh^2m\phi}}\right]
^\frac{\alpha^2}{2m^2}\nonumber \\ &
&\cdot\exp\left[\frac{\sqrt{|f|m^2w^2}\alpha^2}{2m^2}{\arctan}\mbox{h}
\frac{\sqrt{|f|m^2w^2}(1-\sinh^2m\phi)}
{2\sqrt{\cosh^4m\phi+fm^2w^2\sinh^2m\phi}}\right]\quad (\mbox{for~}f<0)
\,,
\end{eqnarray}

\begin{figure}[ht]
\centering
\includegraphics
{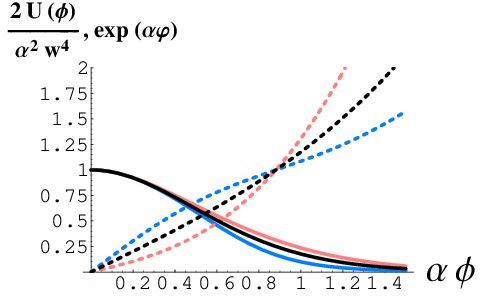}
\caption{The shape of the potential and the development of the volume of the
space. The parameters are chosen as $b=1$ and $m=\alpha$ in the prepotential. 
The solid lines indicate the $\frac{2U(\phi)}{\alpha^2w^4}$, while the broken
lines indicate $e^{\alpha\varphi}(=a^{D-1})$ in arbitrary scales. The color
corresponds to the value of $f\alpha^2w^4$: the black lines correspond to
$f\alpha^2w^4\rightarrow 0$, the magenta lines correspond to $f\alpha^2w^4=7$, and
the cyan lines correspond to $f\alpha^2w^4=-3$.}
\label{fig1}
\end{figure}

In Fig.~\ref{fig1}, the shape of the scalar potential $U(\phi)=
V(\phi)+\frac{1}{f}$ and the development of the volume of the space for the case
with
$b=1$ and
$m=\alpha$. From this figure, we can see that when the constant $f$ is positive,
power-law inflation (for large
$\phi$) is more effective, and when
$f$ is negative, de Sitter expansion (for small $\phi$) is more effective.
Of course, from a cosmological perspective, a precise analysis that takes into
account cosmic time is necessary, but we believe that such a tendency can
be read. Furthermore, since this model ignores the scalar field dependence
 of $f$, more detailed investigation is necessary in the future.


For the case with constant $f$, the slow-roll parameters are expressed by the
single-field prepotential $W(\phi)$, as
\begin{eqnarray}
\varepsilon_H&=&-\frac{\dot{H}}{H^2}=\frac{2(D-2)}{\sqrt{1+fW'(\phi)^2}}\frac{W'(\phi)^2}{W(\phi)^2}\,,\quad
\\
\eta_H&\approx&-\frac{\dot{u}}{Hu}
=\frac{2(D-2)}{\sqrt{1+fW'(\phi)^2}}\frac{W''(\phi)}{W(\phi)}\,,
\end{eqnarray}
where $u=\frac{\dot{\phi}}{\sqrt{1-f\dot{\phi}^2}}$, and the equation of motion
(\ref{eomp}) reduces to $\dot{u}+(D-1)Hu+V'(\phi)=0$.
Thus, the slow-roll condition $\varepsilon_H\ll 1$ is satisfied if
$m\phi\ll 1$ and/or $bm\ll 1$ and/or $\sqrt{1+fW'(\phi)^2}\gg1$.
Thus, in the example of the present section, the power-law inflation at
$u\ne 0$ tends to satisfy the slow-roll condition, if $f>0$. When $f<0$
the opposite tendency occurs. This provides a qualitative explanation for what
depicted in Fig.~\ref{fig1}.

\section{Summary and outlook}
\label{conclusion}

In this paper, we adopted the quantum version of the BPS equation for a class of
gravitating scalar theories, and determined the wave function of the
Universe. A very plausible solution that leads to a probability
density for scalar fields was found. Instead of choosing the initial conditions of
the wave function in the Wheeler--DeWitt equation, the solution is constrained by
adopting the BPS equation. In this study, we found that the initial
probable value of the scalar fields is given by the maximum of the positive
prepotential
$W(\phi)$ and the scale factor of the flat space is zero, for an example.
In the present analysis, we left aside the problem of operator ordering
\cite{HP,TW,Halliwell2,KW,SH,HC} as well as the positivity of the probability
measure.%
\footnote{The most gentle solution to the interpretation would be to define
the expectation value of $\phi^a$ as
$\langle\phi^a\rangle\equiv\frac{\int\phi^a\rho
d^N\phi}{\int\rho d^N\phi}$.} These are issues that remain to be addressed in the
future.

Russo \cite{Russo2} also considered the general nonlinear sigma model. 
In that case, further study is required, since we claim that the probability
measure of scalar fields is proportional to $\sqrt{G}$.
Moreover,
an interesting future challenge will be to search for a model that fits
cosmology within that category.

It is essential to pursue the symmetry that is the origin of the potential shape
of the scalar model. Future work should include considering supersymmetry or
pseudo-supersymmetry, as well as connections with known supergravity models and
theories inspired by string theory.
Further, it is known that string theory compactifications 
naturally lead to slowly varying positive potentials (see \cite{HVWW}, and
references therein) as in the examples in the present paper, so it would be
interesting to explore the connection of our models to string theories.

Furthermore, in determining the restrictions to the quantum BPS equation, the
process of supersymmetrization of the Wheeler--DeWitt equation itself is
important \cite{Graham1,Graham2,OSB,OPR,RB,Moniz}. In that case, a fermionic wave
function will also appear. In such a system, it seems that ideas like third
quantization
\cite{SC,KAHS2,BW} of both bosonic and fermionic wave functions in minisuperspace
will become natural, so research in that direction will likely be deepened in the
future.

\bibliographystyle{apsrev4-1}


\end{document}